\documentclass[final]{aipproc}
\layoutstyle{8x11double}
\def\apj{Astrophysical Journal}
\def\etal{et al}

\def\nat{Nature}
\def\aap{A\&A}

\def\aaps{A\&A Supp. Series}

\def\araa{ARA\&A}
\def\mnras{MNRAS}

\def\apjl{ApJL}

\begin{document}

\title{XTRA: The fast X-ray timing detector on XEUS}
\author{Didier Barret\footnote{On behalf of all of those who were supporting the
proposal to add a fast timing detector in the XEUS focal plane (see
Acknowledgments).}}{ address={Centre d'Etude Spatiale des Rayonnements, \\
9 Avenue du Colonel Roche, 
31028 Toulouse, Cedex 04, France {\it (Didier.Barret@cesr.fr)}}}

\begin{abstract} The Rossi X-ray Timing Explorer (RXTE) has demonstrated that
the dynamical variation of the X-ray emission from accreting neutron stars and
stellar mass black holes is a powerful probe of their strong gravitational
fields. At the same time, the X-ray burst oscillations at the neutron star spin
frequency have been used to set important constraints on the mass and radius of
neutron stars, hence on the equation of state of their high density cores. The
X-ray Evolving Universe Spectroscopy mission (XEUS), the potential follow-on
mission to XMM$-$Newton, will have a mirror aperture more than ten times larger
than the effective area of the RXTE proportional counter array (PCA). Combined
with a small dedicated fast X-ray timing detector in the focal plane (XTRA: XEUS
Timing for Relativistic Astrophysics), this collecting area will provide a leap
in timing sensitivity by more than one order of magnitude over the PCA for
bright sources, and will open a brand new window on faint X-ray sources, owing
to the negligible detector background. The use of advanced Silicon drift
chambers will further improve the energy resolution by a factor of $\sim 6$ over
the PCA, so that spectroscopic diagnostics of the strong field region, such as
the relativistically broadened Iron line, will become exploitable. By combining
fast X-ray timing  and spectroscopy, XTRA will thus provide the first real
opportunity to test general relativity in the strong gravity field regime and to
constrain with unprecedented accuracy the equation of state of matter at
supranuclear density. 
\end{abstract}
\maketitle
\section{Introduction}

The X-rays generated in the inner accretion flows around black holes and neutron
stars carry information about regions of strongly curved spacetime. This is a
regime in which there are important predictions of general relativity still to
be tested, such as the existence of an innermost stable circular orbit. X-ray
spectroscopy and fast timing studies can both be used to diagnose the orbital
motion of the accreting matter in the immediate vicinity of the collapsed star,
where the effects of strong gravity become important. 

With the discovery of millisecond aperiodic X-ray time variability from
accreting black hole and neutron star X-ray binaries (quasi-periodic
oscillations: QPOs), and brightness burst oscillations in neutron stars, RXTE
(\cite{bradt93}) has clearly demonstrated that fast X-ray timing has the
potential to constrain the mass and radius of neutron stars and measure the
motion of matter in strong gravity fields (see \cite{klis00} and e.g. Lamb these
proceedings). As pointed out by van der Klis, it is now time to turn these
diagnostics into true tests of general relativity. For this, a follow-up to the
RXTE Proportional Counter Array (PCA) should have a collecting area of at least
ten times larger (more than 6 m$^2$) and an energy resolution at least five
times better (i.e. $\sim 200$ eV at the Iron line).
 
There are two possible implementations for such an instrument. One is
collimated, in which the collecting area is the same as the detector area (as in
the PCA), and one in which the two are decoupled through the use of focussing
optics. In the first case, beside proportional counters (e.g. Zhang, these
proceedings), large area thick Silicon PIN diodes could be considered (e.g.
\cite{barret01,kaaret01} and Kaaret, these proceedings), although it is unclear
whether they will provide an improvement in energy resolution and whether the
background could be kept as low as in PCA-like proportional counters. On the
other hand, with the use of focussing optics, the immediate advantage is that
the detector can be made very small with low background and good energy
resolution, allowing timing and spectroscopy to be carried out simultaneously.

This paper describes the XEUS Timing for Relativistic Astrophysics (XTRA)
instrument considered for the focal plane instrumentation of the XEUS mission
\cite{barret03}. Combining the huge collecting area (30 m$^2$) and broad band
pass (up to $\sim 30$ keV) of the XEUS mirror optics with Silicon drift
chambers, XTRA will fulfill the two main requirements for a follow-up to the
PCA. The science case for such an instrument was presented by van der Klis in
the context of XEUS \cite{klis03}, and will be emphasized again in many papers
in these proceedings (see the contributions by Kaaret, van der Klis, Lamb,
Miller, Psaltis, Strohmayer, \ldots). It will not be repeated here. Instead in
this paper, first I will briefly summarize the current XEUS mission profile,
then I will present some sensitivity estimates for XTRA and show the results of
some simulations of QPOs and burst oscillations, using the expected count rates
for XTRA. Finally, I will describe the detector implementation (see also
\cite{barret03xtra,staubert03}).

\section{XEUS in brief} XEUS: The X-ray Evolving Universe Spectroscopy mission
represents a potential follow-on mission to the ESA XMM$-$Newton cornerstone
observatory currently in orbit. The XEUS mission was considered as part of the
ESA Horizon 2000 plus program within the context of the International Space
Station (ISS). XEUS is the next logical step forward in X-ray astrophysics after
the current set of great observatories (XMM$-$Newton and Chandra), have
completed their operational lives. The scientific objectives of XEUS are however
so demanding that the mission will clearly represent a major technological
challenge compared to past astrophysics missions. Its development and ultimate
success rely on the capability to achieve a key breakthrough in the size of an
optic capable of entering orbit.

The primary aim of XEUS is the astrophysics of the most distant discrete objects
in the Universe (see \cite{hasi00} for details). The specific scientific issues,
which XEUS will address can be summarized as follows: 1) To measure the spectra
of objects with a redshift z$>4$ at flux levels below 10$^{-18}$ ergs cm$^{-2}$
s$^{-1}$, which is just about 1000 times fainter than XMM$-$Newton 2) To
determine from the X-ray spectral lines the redshift and thus age of these very
faint objects that may not have easily identified optical counterparts 3) To
establish the cosmological evolution of matter in the early Universe through the
very clear means of the study of heavy element abundances as a function of
redshift, i.e. the role of element evolution as the Universe aged through galaxy
formation in the associated early stellar processes.

\begin{figure}[!t]
\includegraphics[height=.31\textheight]{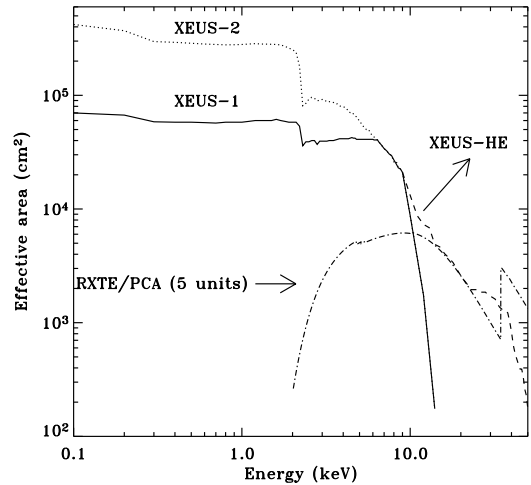}
\caption{Comparison between the XEUS$-$1, XEUS$-$2 and PCA (5 units) effective
areas. The proposed High Energy extension (XEUS$-$HE) for the mirrors is also
represented (data taken from the Telescope Working Report \cite{asch01}).}
\label{barretd_f1}
\end{figure}

To meet these scientific objectives, XEUS consists of two free flying
satellites: a mirror spacecraft (MSC) and a detector spacecraft (DSC) separated
by 50 meters and aligned in orbit by an active orbital control and alignment
system \cite{bavdaz03,parmar03}. In the baseline mission scenario, the required
large aperture mirrors (see Figure \ref{barretd_f1}) cannot be deployed in a
single launch, making XEUS a two step mission, in which the growing of the
mirrors involves the facilities in place on the ISS. However, given the
uncertainties related to the future of the ISS, alternative mission scenarios
are currently being considered at ESA. 

In the baseline, XEUS$-$1 comprising the mated mirror and detector satellites
will be launched by an Ariane 5 type of launcher in a low earth orbit, with an
inclination similar to the ISS. After having completed a few years of
observations, MSC1 will go docking on the ISS, where the external mirror
segments are waiting. Using the robotic arms on the ISS, the mirror segments
will be added to MSC1 to reach the final configuration of MSC2. A second
detector spacecraft will then be launched to operate at the focus of MSC2:
XEUS$-$2 will then be born. The effective areas of XEUS$-$1 and XEUS$-$2
\cite{asch01} are shown in Figure \ref{barretd_f1}, and compared with the
effective area of the RXTE/PCA (5 units combined). The proposed high energy
extension in which the inner mirror shells are coated with supermirrors is also
shown. This yields $\sim 20000$ cm$^2$ at $\sim 9$ keV and still $\sim 1700$
cm$^2$ at 30 keV.

Although XEUS was designed to explore the most distant regions of the Universe,
we now show how such an ambitious mission also has great potential for studying
the brightest sources in the sky.
\section{XTRA sensitivity for timing studies} For the sensitivity computations,
we have assumed that the timing detector is made of 300 microns of Silicon lying
above 2 mm of CdZnTe (see below). This produces a more or less flat detector
response up to 80 keV. Table \ref{barret_t1} gives the count rates expected from
some sources, accounting for typical source spectra and interstellar absorption
(for galactic sources, this suppresses most of the photons below $\sim 1$ keV).
The Crab produces about 14000 counts/s in the PCA, which is just about 17 times
less than for XEUS$-$1 and 57 times less than for XEUS$-$2 (larger conversion
factors apply for spectra softer than the Crab spectrum; 20 is a good average).

\begin{table}[!t]  
\label{barret_t1}
\begin{tabular}[h]{lccc}
\hline Source name & XEUS$-$1& XEUS$-$2 & C$_{\rm E >10 keV}$ \\
\hline Crab & 250000 & 800000 & 5000 \\
Sco X$-$1 & 1200000 & 3800000 & 10000 \\
X-ray burst & 120000 & 220000 & 2000 \\
SAXJ1808$-$3659 & 30000 & 130000 & 300 \\
\hline
\end{tabular}
\caption{Examples of total count rates above 0.5 keV and above 10 keV (C$_{\rm E
>10 keV}$) in counts/s. The spectrum of Sco X$-$1 which is variable corresponds
to 60000 counts/s in the RXTE/PCA (2.5--30 keV). The X-ray burst input spectrum
is a blackbody of 1.5 keV yielding an Eddington luminosity at 8.5 kpc.
SAXJ1808$-$3659 is the millisecond pulsar taken at the peak of its 1996
outburst. }

\end{table}
\subsection{QPO detection sensitivity} Let us now compute the sensitivity for
QPO detection. The signal to noise ratio $n_\sigma$ at which a QPO is detected
in a photon counting experiment is approximately: $$
n_\sigma = {1\over2}{S^2\over B+S}r_S^2\left( T\over\Delta\nu
\right)^{1/2}$$
where $S$ and $B$ are the source and background count rates, respectively, $r_S$
is the (RMS) root mean squared amplitude of the variability expressed as a
fraction of $S$, $T$ the integration time and $\Delta\nu$ the bandwidth of the
variability. The coherence time of the signal is related to the width of the QPO
as $\tau=1/\pi\Delta\nu$.

\begin{figure}[!t]
\includegraphics[height=.3\textheight]{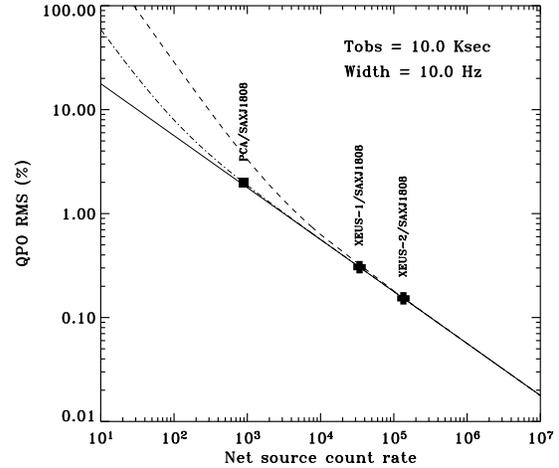}
\caption{Comparison between the XTRA (solid line), RXTE/PCA (dot-dashed line)
and EXTRA (dashed line) sensitivities for QPO detection ($5\sigma$ in 10 ksec,
signal width 10 Hz). An illustrative example is provided by the millisecond
pulsar. As can be seen, a factor of $\sim 10$ improvement in sensitivity over
the RXTE/PCA is obtained with XEUS/XTRA.}
\label{barretd_f2}
\end{figure}

From the above formulae, assuming B $\sim 0$ appropriate for XTRA, one can
estimate the RMS amplitude corresponding to a $5\sigma$ QPO detection as a
function of the source count rate (Figure \ref{barretd_f2}). One can also do the
same computations for the PCA (assuming a background of 100 counts/s) and for an
EXTRA-like instrument of the type proposed to ESA in response to the call for
F2/F3 mission proposals \cite{barret01}. EXTRA was designed as a collimated
instrument with a detector made of large area Si PIN diodes, covering a total
effective area of 6.7 m$^2$ (see also Kaaret, these proceedings for a similar
design). Estimated background rate for such an instrument in a low earth orbit
gave about 2500 count/s. Figure \ref{barretd_f2} shows that i) XTRA provides
better than one order of magnitude sensitivity improvement in terms of RMS for
QPO detections over the PCA ii) thanks to the negligible background, XTRA will
enable the timing of very faint X-ray sources (down to less than 0.1 mCrab; e.g.
binaries in external galaxies), whereas the background of an EXTRA like mission
is of the order of 5 mCrab.

Given the scaling of the above formula, and the increase of count rates between
the PCA and XTRA, a QPO detected at $5\sigma$ with the PCA will be detected at
$\sim 100\sigma$ with XEUS$-$1. Similarly, XTRA on XEUS$-$1 will detect signals
at the same level of significance as the PCA but for an observing time $\sim
400$ times shorter, and for the strongest signals over tens of milliseconds.
XTRA will thus enable us to detect kilo-Hz QPOs at much lower amplitudes,
especially in the domain in which the PCA loses the signal towards the highest
frequencies \cite{klis00}. This is a region where we might be able to see the
saturation of the frequency at the innermost stable circular orbit (which gives
a measure of the neutron star mass), if the orbital interpretation for the
signal is correct. One might also detect QPOs at frequency above 1330 Hz which
is the maximum observed today \cite{vanstraaten00}. As stressed by Miller in
these proceedings, detection of QPOs at frequencies as high as 1800 Hz would
allow us to eliminate all standard nucleonic or hybrid quark matter equations of
state, leaving only strange stars. 

\begin{figure}[!ht]
\includegraphics[height=.31\textheight]{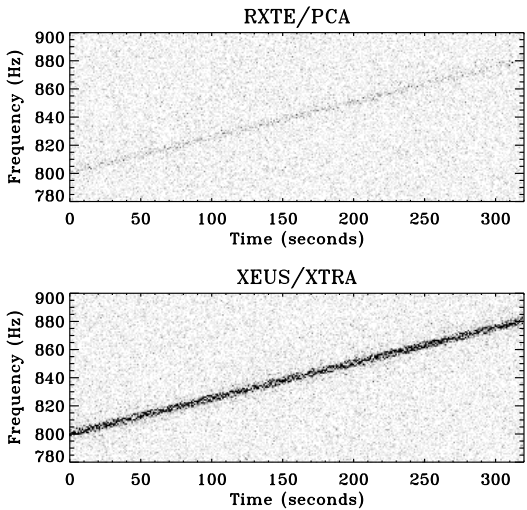}
\caption{{\it Top:} Dynamical power density spectrum of a source producing 3000
counts/s in the RXTE/PCA and a QPO of RMS $\sim 6.5$\%. The QPO is produced by a
pulsar like signal (i.e. the signal is always there) and the phase is changed
randomly after 200 cycles (its width is about 4 Hz, FWHM). The power spectra are
integrated over a 1 second time interval. The frequency of the QPO increases
linearly with time at a rate of 0.25 Hz/s. {\it Bottom:} The same simulations
but for XTRA with a count rate 15 times larger than the PCA one. Not only the
QPO can be detected in all 1 second intervals at the expected frequency, but its
width can also be inferred.}
\label{barretd_f3}
\end{figure}

As stressed above, QPOs will be detected on very short timescales, and depending
on the nature of the signal, possibly within their coherence times. This would
by itself allow us to discriminate between the various classes of models
\cite{klis03}, yielding in some cases constraints on the compact object mass,
angular momentum and orbital radius at which the QPOs are produced (see e.g.
Abramowicz \& Kluzniak, Miller these proceedings). For illustrative purposes, we
have first carried out a simulation of a QPO, which is a pulsar like signal of
constant amplitude (6.5\% RMS for a source count rate of 3000 count/s in the
PCA) for which the phase is changed randomly every 200 cycles. The frequency of
the signal shifts at a constant rate of 0.25 Hertz per second. This produces a
QPO of width $\sim 4$ Hz corresponding to a quality factor Q$=\nu / \Delta \nu
\sim 200$. Q values as large as 200 have already been reported for QPOs measured
in power density spectra integrated over 100 seconds or more (e.g.
\cite{berger96} for 4U1608$-$52). Here we have computed power density spectra
over 1 second time intervals, and produced a dynamical power density spectrum
for both the PCA and XTRA, assuming a count rate of 45000 counts/s for the
latter. The result is shown in Figure \ref{barretd_f3}. It clearly shows that,
although the frequency evolution of the QPO can be tracked in the PCA image,
only XTRA can measure its frequency and width on short time scales. 

\begin{figure}[!t]
\includegraphics[height=.31\textheight]{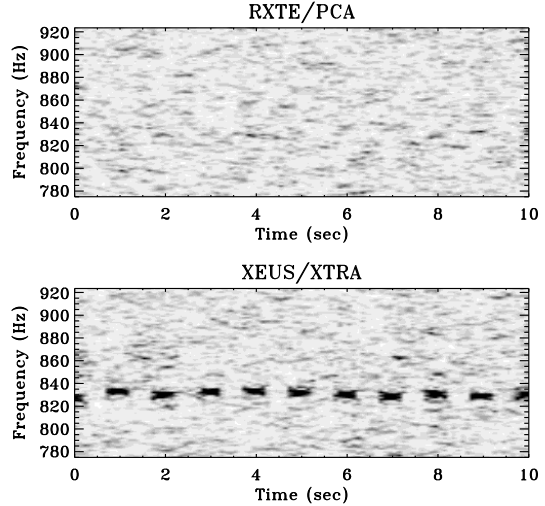}
\caption{Time-frequency image made by stacking Rayleigh periodograms computed on
0.4 second every 0.1 second. The QPO is a succession of finite lifetime
wavetrains of $\sim 0.25$ second duration (200 cycles at 830 Hz). The QPO RMS is
6.5\%. There is only one wavetrain per second. The source count rate is 3000
counts/s in the RXTE/PCA ({\it top}) and 45000 counts/s in the XEUS/XTRA ({\it
bottom}). The bold regions in the XTRA image coincide with the appearance of the
QPO wavetrain. The signal is hardly visible in the PCA image}
\label{barretd_f4}
\end{figure}

We have carried out a second set of illustrative simulations to demonstrate that
under somewhat favorable conditions, one can localize the signal directly in the
time domain. This time the QPO is made of a succession of wavetrains of finite
lifetime (QPO at 830 Hz keeping the same frequency for 200 cycles, the RMS of
the QPO is again fixed at 6.5\% for a source at 3000 count/s in the PCA and
there is one wavetrain per second). To explore the signal in the time domain, we
slide a window of 0.4 second width over the data with a time step of 0.1 second.
Then we search for a periodic signal using a standard Rayleigh test. The
periodograms (2.5 Hz of resolution) so obtained are stacked together to form a
time-frequency image. One expects that when the window is centered on the QPO
wavetrains, the periodogram will reach a maximum at the QPO frequency, and this
will appear as a strong excess in the image. A comparison between a
time-frequency image obtained for PCA and XTRA count rates is shown in Figure
\ref{barretd_f4}. Whereas it is impossible to locate the signal in the PCA
image, it stands out very clearly in the XTRA image. Once located in the time
domain, one can look at the time profile of the oscillations (in the present
simulation folding the data produces a nice sinusoid) and study what causes the
signal to lose its coherence. One can also do waveform fitting, correlated
timing and spectral analysis and so on. With sophisticated time-frequency
analysis tools (singular spectrum analysis, maximum enthropy methods, \ldots),
even under less favorable conditions (i.e. shorter coherence times, rapid
frequency drifts, superposition of wavetrains, \ldots), with XTRA it will be
possible to investigate the true nature of the signal by looking for the first
time at its properties in the time domain.

\begin{figure}[!ht]
\includegraphics[height=.31\textheight]{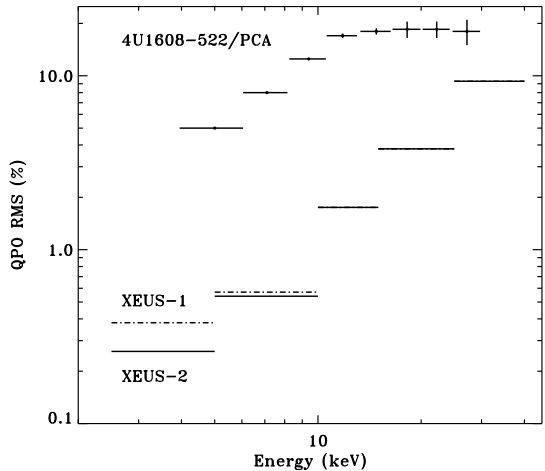}
\caption{Sensitivity for QPO detection (RMS \%) as a function of energy
(5$\sigma$, 10 kseconds). The estimates are compared to the RXTE/PCA data from
the neutron star low-mass X-ray binary 4U1608$-$52 (data taken from
\cite{berger96}).}
\label{barretd_f5}
\end{figure}

In general, when plotting the QPO RMS as a function of energy, it is found that
it increases, to reach a saturation around 10$-$20 keV (e.g. Figure
\ref{barretd_f5} from \cite{berger96}). One limitation of the use of focussing
optics is their energy band pass which drops off quickly after 10 keV (see
Figure \ref{barretd_f1}). In the case of XEUS however, the use of advanced super
mirrors extends the energy response up to 30 keV. In fact, the effective area of
the XEUS mirrors is comparable or even larger than the PCA (5 units) up to 30
keV. In the current detector implementation for XTRA, the use of CdZnTe below
the Silicon would match the high energy response of the mirrors. One can compute
the limiting RMS (using equation above), as a function of energy for a given
source energy spectrum. This is shown in Figure \ref{barretd_f5} and compared
with the observed data points for the neutron star X-ray binary 4U1608$-$52
\cite{berger96}. This figure shows that thanks to the high energy extension of
the mirrors, XTRA will be more sensitive than the PCA over the whole energy
range QPOs were detected.

\subsection{Coherent signal detection sensitivity} 
\begin{figure}[!ht]
\includegraphics[height=.32\textheight]{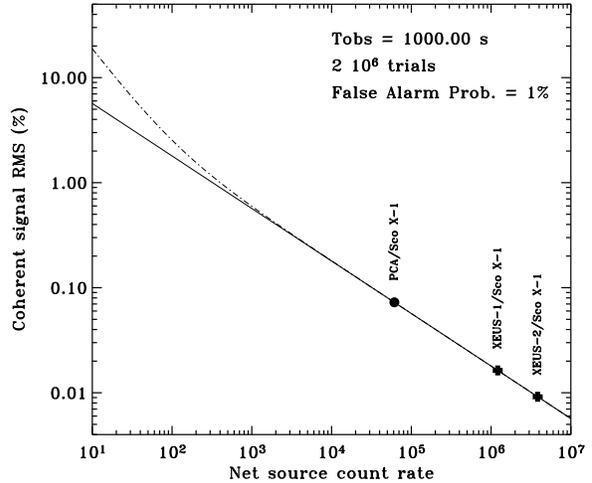}
\caption{Comparison between the XEUS (solid line), RXTE/PCA (dot-dashed line)
sensitivities for coherent signal detection (1 ksec). The detection level
corresponds to a false alarm probability of 1\% for $2\times 10^6$ trials. So
far, no pulsations have been detected from Sco X$-$1.  The XEUS$-$1/XTRA
sensitivity is 10 times better than the current RXTE/PCA sensitivity, and
failure to detect pulsations at this level would demand major revision of our
current ideas about low-mass X-ray binaries.}
\label{barretd_f6}
\end{figure}

For a coherent signal ($T$$
<$$
1/\Delta\nu$), the familiar exponential detection regime applies, with
false-alarm probability $\sim$$
\exp[-{S^2r_S^2T/2(B+S)}]$. One can compute the RMS for the detection of a
coherent signal at a given false alarm probability (Figure \ref{barretd_f6}).
The figure shows that XTRA would be able to detect pulsations at the 0.01\% RMS
level in Sco X$-$1. Why so far no classical low-mass X-ray binaries, such as Sco
X$-$1, have shown pulsations in their persistent emission remains very puzzling.
Several competing explanations have been put forward, as for example that they
may be rotating at sub-millisecond periods or may contain compact neutron stars
\cite{klis03}. Thanks to its improved sensitivity for periodic signals XTRA will
enable us to test these hypothesis, which have their own  constraints on the
equation of state of dense matter. 

In addition, it has been recently suggested that the neutron star spin
frequencies inferred from burst oscillations have an upper limit of $\sim$750~Hz
\cite{chakra03}.  If this is because of angular momentum removed by
gravitational radiation, then searches for periodic gravitational  waves emitted
by these systems require highly accurate measurements of their spin periods. 
XTRA could naturally achieve this task.

\begin{figure}[!ht]
\includegraphics[height=.31\textheight]{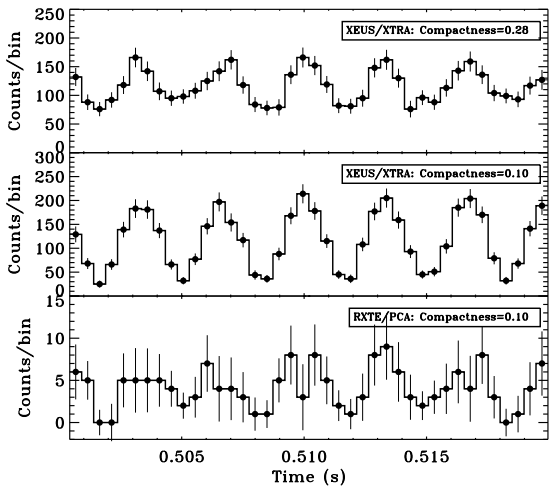}
\caption{Simulated light curve of a burst oscillation at 300 Hz as seen by the
RXTE/PCA and XEUS/XTRA. The burst produces 15000 counts/s in the PCA, and 400000
count/s in XTRA. The simulations take into account gravitational light bending
in a  Schwarzschild spacetime following the method described in \cite{nath02}
(one spot and a cosine emission diagram). For XTRA, the light curve is computed
for two neutron star compactnesses (M/R=0.1 and 0.28). With XTRA the
oscillations are directly visible in the light curve even at the highest
compactness, whereas they are invisible in the PCA data. With XTRA, which will
also be sensitive to the harmonic content of the signal, beside the stellar
compactness, both the hot spot and viewing geometries could be constrained.}
\label{barretd_f7}
\end{figure}

Concerning burst oscillations, XTRA with its optimum band pass for X-ray bursts,
will allow the oscillations to be detected within one cycle (see Figure
\ref{barretd_f7}). These oscillations are probably caused by rotational
modulation of a hot spot on the stellar surface. The emission from the hot spot
is affected by Doppler boosting, relativistic aberration and gravitational light
bending (whose magnitude increases with the neutron star compactness, see Figure
\ref{barretd_f7}). By fitting the waveform, it will be possible to investigate
the spacetime around the neutron star, and simultaneously constrain its mass and
radius, and hence determine the equation of state of its high density core (see
Strohmayer, these proceedings, but see also Poutanen for an example of what can
be done by studying the energy dependence of pulse profiles of accreting
millisecond pulsars).

\section{Detector implementation} To meet the scientific objectives of XTRA, its
detector must be capable of handling up to 3 Mcts/s (XEUS$-$1) and 10 Mcts/s
(XEUS$-$2) (equivalent to a 10 Crab source, see Table 1) with a timing
resolution of $\sim 10
\mu$s and a deadtime as low as possible, with the requirement that the latter
must be measured accurately. In addition, the detector energy range should match
closely the high energy response of the mirrors.

\subsection{Silicon Drift Detector} 
\begin{figure}[!t]
\includegraphics[height=.175\textheight]{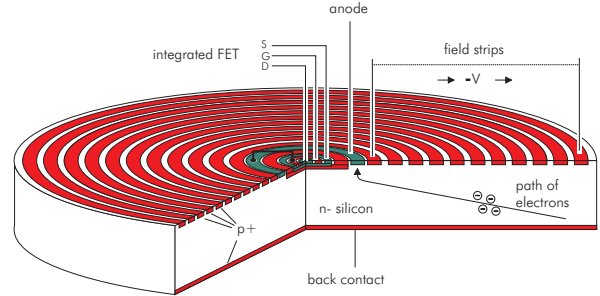}
\caption{Schematic cross section of a cylindrical Silicon Drift Detector (SDD).
Electrons are guided by an electric field towards the small collecting anode
located at the center of the device. The first transistor of the amplifying
electronics is integrated on the detector ship (drawing kindly provided by P.
Lechner).}  \label{barretd_f8}
\end{figure}

Among the fast X-ray detectors currently
available, Silicon Drift Detectors (SDDs) are the most promising
\cite{stru00,lech01}.  The SDD is a completely depleted volume of Silicon in
which an arrangement of increasingly negative biased rings drive the electrons
generated by the impact of ionising radiation towards a small readout node in
the center of the device. The time needed for the electrons to drift is much
less than 1 $\mu$s.  The main advantage of SDDs over conventional PIN diodes is
the small physical size and consequently the small capacitance of the anode,
which translates to a capability to handle high count rates simultaneously with
good energy resolution. To take full advantage of the small capacitance, the
first transistor of the amplifying electronics is integrated on the detector
chip (see Figure
\ref{barretd_f8}). The stray capacitance of the interconnection between the
detector and amplifier is thus minimized, and furthermore the system becomes
practically insensitive to mechanical vibrations and electronic pickup.

\begin{figure}[!ht]
\includegraphics[height=.225\textheight]{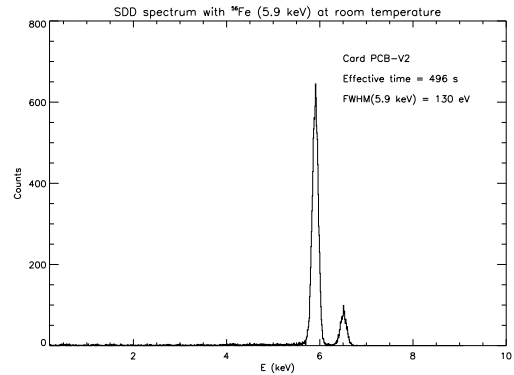}
\caption{Count spectrum of a 5mm$^2$ Silicon Drift Detector as measured at room
temperature at CESR. The energy resolution reached is comparable to CCD
resolution. Such a resolution would allow spectroscopy to be combined with
timing.}  \label{barretd_f9} 
\end{figure}

A spectrum of a 5 mm$^2$ SDD, with a customized amplifying electronics, is shown
in Figure \ref{barretd_f9} to illustrate that energy resolution as good as 130
eV can be achieved, even at room temperature. Energy resolution of better than
$\sim 200$ eV can be maintained for count rates up to 10$^5$ cts/s (e.g.
\cite{lech01}).

With such an energy resolution (comparable to what CCDs are currently
providing), spectroscopy and timing can be used as independent tools to probe
the strong gravity regions around compact objects. A particular emphasis should
be put on the relativistically broadened Fe K$\alpha$ lines, recently detected
by XMM$-$Newton and Chandra (e.g. \cite{miller03}). The line profile which is
affected by several physical processes (Doppler shifts, beaming, gravitational
redshifts) gives complementary information about the same region, as explored
with millisecond timing. The lines are relatively broad (1 keV or more) so that
the 200 eV energy resolution of XTRA is sufficient. This again offers
independent opportunities to constrain the compact object mass and spin, as well
as the radius at which the line is produced. Similarly, as pointed out by
Strohmayer (these proceedings) lines in X-ray bursts \cite{cottam} might be
detectable, even with a 200 eV spectral resolution. Measurement of a
gravitational redshift yields a direct estimate of the compactness of the
neutron star, and yet another constraint on the equation of state of dense
matter (see Strohmayer, these proceedings).

Finally with a rather conservative low energy threshold $\sim 1$ keV, XTRA will
be very sensitive to very soft sources (e.g. quiescent neutron star transients,
supersoft sources, millisecond puslars, isolated neutron stars \ldots), in an
energy domain never explored before (2.5 keV is the current energy threshold of
the RXTE/PCA).
 
\subsection{The XTRA detector} For timing studies, deadtime is always a critical
issue. Deadtime will include contributions from the signal rise time, the charge
sensitive amplifier, and the shaping amplifier. The first two of these can be
very short, and the limiting contribution is that of the amplifier, where a
trade-off between speed and energy resolution is necessary. Shaping time
constants as short as 50 nanoseconds (ns) have been found to be usable
\cite{stru00}. The challenging requirement for XTRA is to achieve an accurately
measured deadtime of $\sim 500$ ns per event. Using currently available devices
and pipelining techniques, the analog-digital conversion stage is not a limiting
factor at these speeds.

\begin{figure}[!t]
\includegraphics[height=.17\textheight]{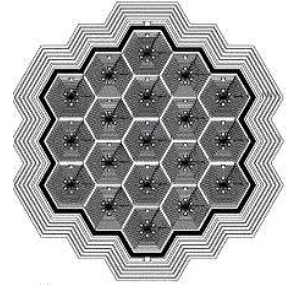}
\caption{SDD array made of 19 hexagon cells of 5 mm$^2$ \cite{lech01}. The
overall size of the detector is just about 1 cm$^2$.} 
\label{barretd_f10} 
\end{figure}

A 500 ns deadtime per event corresponds to a 5\% deadtime for a source producing
$10^5$ cps/s. To handle $10^6$ cps/s with a reasonable deadtime, one must
therefore distribute the focal beam over $\sim 10$ detectors or more. The best
and easiest solution could thus be a detector made of an ensemble of about $\sim
10$ separate SDDs on a single wafer. Such SDD arrays already exist, as shown in
Figure \ref{barretd_f10}. {\it This detector should therefore be operated out of
focus}. For XEUS$-$1, the out of focus distance is of the order of 10 cm. This
could be accomplished either by a mechanical construction, or by changing the
distance between the mirror and detector satellites. Although this will require
a careful study, both solutions appear to be feasible within the current XEUS
mission design. The requirements in terms of real estate on the detector
spacecraft are not constraining, in particular because no complicated cooling
systems will be necessary. 

As mentioned above, a high energy extension (above 10 keV) is proposed for the
mirrors \cite{asch01}. SDDs are currently produced with thicknesses of 300
microns, which is adequate to cover the energy range below 10 keV. Although
there are on-going efforts to make thicker devices, the best match of the high
energy response of the mirrors will require the SDD array to be associated with
a higher density detector located underneath. Among the potential high energy
semi-conductor detectors, CdZnTe
\cite{budt01} stands today as a very promising solution. Such a detector would
both ensure the overlap in energy range with the SDD array, and provide a flat
energy response up to $\sim 80$ keV and 10 microsecond timing resolution
\cite{budt01}. Count rates should not exceed a few thousands counts/s in the
CdZnTe detector.

The goal for XTRA is to send to the ground the time and energy information of
every photon. For most sources, data compression will make this possible (within
a 2 Mbits/s data rate) without compromising either time or energy resolution.
For the very brightest sources, this can still be done with a restricted number
of energy channels. However, within the timeframe of XEUS, on-board memory and
telemetry rate should not be considered as real issues.

The detector will be exposed to high radiation doses during operations and one
must therefore consider its radiation hardness. The main limitation in the
maximum acceptable dose arises from the JFET connected to the collecting anode
on the back of the device \cite{leutenegger00}. High energy photons absorbed in
the transistor region increase the amount of oxyde charge and interface traps,
thus reducing the charge carrier lifetimes, and thus contributing to increase
the leakage current. Laboratory measurements indicate however that a 300 micron
thick SDD survives a radiation dose of $\sim 10^{13}$ incoming high energy
photons (E$>12$ keV) \cite{leutenegger00}. This is  equivalent to a continuous
exposition of 3 years at 10$^5$ photons/s. Concerning particles, the
XMM$-$Newton EPIC pn cameras, which have similar detector technology is
performing extremely well in space \cite{struder01AA}. So, the device selected
can be clearly considered as radiation hard.

To get ready when XEUS gets approved in the ESA science program, we have
proposed to the French Space Agency an R\&T program, aimed at characterizing the
performance of the detector for XTRA and starting to work on deadtime issues and
on-board data processing. The proposal was approved in 2002 (before CNES
experienced internal problems). Hopefully funding should arrive in 2004. This
hardware program will be carried out in close collaboration with MPE/MPI,
University of T\"ubingen and the ROENTEC company which has a great expertise on
high count rate electronics for Silicon drift detectors.

\section{Conclusions} 
The XEUS Timing for Relativistic Astrophysics instrument, by combining the huge
aperture of the XEUS mirrors with a silicon drift detector array in the focal
plane meets the two most important science requirements for a follow-up to the
RXTE/PCA. It will have a collecting area more than ten times larger than the one
of the PCA, and a much improved energy resolution. Through fast X-ray timing and
spectroscopy, XTRA will thus provide the first opportunity to test general
relativity in the strong gravity field regime and constrain with unprecedented
accuracy the equation of state of dense matter.
\begin{theacknowledgments} I am extremely grateful to the following colleagues
who gave their support to the proposal for a fast X-ray timing detector in XEUS:
J.L. Atteia, T. Belloni, H. Bradt, L. Burderi, S. Campana, A. Castro-Tirado, D.
Chakrabarty, P. Charles, S. Collin, S. Corbel, C. Done, G. Dubus, M. Gierlinski,
J. Grindlay, A. Fabian, R. Fender, E. Gourgoulhon, J.M. Hameury, C. Hellier, E.
Kendziorra, W. Kluzniak, E. Kuulkers, S. Larsson, J.P. Lasota, T. Maccarone, D.
de Martino, K. Menou, C. Miller, F. Mirabel, M. Nowak, J.F. Olive, S. Paltani,
R. Remillard, J. Rodriguez, R. Rothschild, T. di Salvo, R. Sunyaev, M. Tagger,
M. Tavani, L. Titarchuk, G. Vedrenne, N.  White, R. Wijnands, J. Wilms, A.
Zdziarski, W. Zhang.

I am also grateful to  G. Hasinger, A. Parmar and M. Bavdaz for helpful
information about the XEUS mission in general. I thank J.F. Olive and J.P.
Chabbert for helping me out to produce some of the simulations presented in this
paper, and M. Ehanno and O. Godet for providing me with the spectrum of the SDD.
Many thanks also to J. Wilms and C. Miller for comments on the manuscript.

\end{theacknowledgments}
\bibliographystyle{aipproc}   

\end{document}